\documentclass[10pt,a4paper,twoside]{article}
\usepackage{epsfig}
\usepackage{baltlat6}
\usepackage{array}
\usepackage{here}
\begin{document}
\ \
\vspace{0.5mm}
\setcounter{page}{1}
\vspace{8mm}

\titlehead{Baltic Astronomy, vol.\,xx, xxx--xxx, 2011}

\titleb{STARK BROADENING OF CARBON AND OXYGEN LINES IN HOT DQ          
WHITE DWARF STARS: RECENT RESULTS AND APPLICATIONS}

\begin{authorl}
\authorb{P. Dufour}{1},
\authorb{N. Ben Nessib}{2},
\authorb{S. Sahal-Br\'echot}{3} and
\authorb{M. S. Dimitrijevi\'c}{3,4}
\end{authorl}

\begin{addressl}
\addressb{1}{D\'epartement de Physique, Universit\'e de Montr\'eal,         
Montr\'eal, H3C 3J7, Canada; dufourpa@astro.umontreal.ca}
\addressb{2}{INSAT (National Institute of Applied Sciences and Technology), University of Carthage, Tunis, Tunisia; nebil.bennessib@planet.tn}
\addressb{3}{Observatoire de Paris, LERMA, CNRS, UMR 8112, 5 Place Jules Janssen, 92190 Meudon, France; sylvie.sahal-brechot@obspm.fr}
\addressb{4}{Astronomical Observatory, Volgina, 7, 11060 Belgrade 38, Serbia; mdimitrijevic@aob.bg.ac.rs}
\end{addressl}

\submitb{Received: 2011 June 27; accepted: 2011 July 14}

\begin{summary} 
  White dwarf stars are traditionally found to have surface
  compositions made primarily of hydrogen or helium. However, a new
  family has recently been uncovered, the so-called Hot DQ white
  dwarfs, which have surface compositions dominated by carbon and
  oxygen with little or no trace of hydrogen and helium (Dufour et
  al. 2007, 2008, 2010). Deriving precise atmospheric parameters for
  these objects (such as the effective temperature and the surface
  gravity) requires detailed modeling of spectral line profiles. Stark
  broadening parameters are of crucial importance in that context. We
  present preliminary results from our new generation of model
  atmosphere including the latest Stark broadening calculations for
  CII lines and discuss the implications as well as future work that
  remains to be done.
\end{summary}


\resthead{Stark broadening of carbon and oxygen lines in Hot DQ white dwarfs}
{Dufour et al.}

\sectionb{1}{INTRODUCTION}

A new spectral class of white dwarf stars with surface compositions
dominated by carbon and oxygen, the Hot DQs, has recently been
uncovered about 4 years ago (Dufour et al. 2007). The first generation
of model atmosphere used to analyze these Hot DQ stars revealed that
they were both hydrogen and helium deficient and that they all
clustered in a very narrow range of effective temperature between
18,000 to 23,000 K (Dufour et al 2008). Follow-up high signal-to-noise
spectroscopic observations of these rare white dwarfs using the MMT
(6.5 m) and the Keck (10 m) telescopes also revealed Zeeman splitted
line profiles in more than half of these stars, indicating the
presence of strong surface magnetic fields in the mega Gauss range
(Dufour et al. 2009, 2010). Luminosity variations have also been
observed in five of the 14 known Hot DQ white dwarfs (Montgomery et
al. 2008, Barlow et al. 2008, Dunlap et al. 2010, Dufour et al. 2011),
opening up a new window through which one may study these stars by
means of asteroseismological studies.

One of our main challenges is now to successfully explain and
understand the extraordinary properties and characteristics of these
stars, as well as the place they occupy in stellar evolution. In order
to achieve this, atmospheric parameters (effective temperature,
surface gravity, surface chemical compositions etc.) must be
determined accurately. Of particular importance is a better
determination of the surface gravity since this would severely
constrain the mass of the progenitors via the initial-final mass
relationship. For example, an extremely high surface gravity may
indicate that these white dwarf stars have evolved from some of the
most massive stars on the main sequence that do not explode as
supernovae. As such, they could have cores made of elements much
heavier than carbon and oxygen.

In the context of carbon/oxygen dominated atmosphere, it is of utmost
importance to have good Stark broadening damping constants in order to
derive meaningful atmospheric parameters from line profiles. However,
Stark damping constants for all of the hundreds of CII lines, from UV
to optical, observed in Hot DQ white dwarfs are not readily available
from the literature (for example, only a few lines can be found in
Griem 1974, Goly and Weniger 1982, Djenize et al. 1988, Blagojevic et
al. 1999, Sreckovic et al. 2000, Mahmoudi et al. 2004 and many other
references, clearly insufficient for our needs). Moreover, since the
first detailed analysis of Hot DQ white dwarfs presented in Dufour et
al. (2007, 2008) were based on low signal-to-noise ratio SDSS
observations, the first generation of Hot DQ model atmosphere were
built simply using the standard approximation (see Castelli 2005):
$\gamma_s/N_e = 10^{-8}n^5_{eff}$ where $\gamma_s$ is the Stark width
of the line in angular frequency units and $n_{eff}$ is the effective
quantum number (Kurucz linelist also gives damping constants for a few
CII lines, but these values are only listed for a single temperature
of 10,000 K and are based on questionable approximations as
well). While this approximation, which is loosely a fit by Peytremann
(1972) to detailed calculations by Sahal-Br\'echot \& Segre (1971),
was enough to reveal the basic properties of Hot DQ stars, it is
certainly not appropriate for a precise determination of atmospheric
parameters, specially the surface gravity, now that high S/N
spectroscopic data are available.

In order to go beyond these limitations from input data in our stellar
modeling, we thus calculated widths and shifts for all the isolated
lines of the CII ion due to electron collisions, recalculated Hot DQ
model atmosphere grids and refitted the CII line profiles
appropriately.

\sectionb{2}{CALCULATIONS AND RESULTS}

Stark broadening parameters were determined within the semi-classical
perturbation method for 1002 CII lines between $\sim$400 and
10,000~\AA~ in the VALD database (see Larbi-Terzi et al. 2010,
Sahal-Br\'echot 2010 and Sahal-Br\'echot et al. 2011). The calculations
were performed for an electron density of $10^{17}$ cm$^{-3}$ and
temperatures between 5,000 K and 100,000 K. In order to facilitate the
inclusion of all these calculations in our stellar atmosphere code, we
fitted, for each CII line, a smooth function of the form $\rm log (w)=
D_1 + D_2~log (T) + D_3~(log (T))^2$, where w is the FWHM width in
angstrom. The D$_i$ for each line can then easily be implemented into
our linelist in order to get the correct width for the temperature and
electron density at each depth of a given model atmosphere. It is
noted that the new Stark widths calculated here are significantly
different than those obtained from the above approximation. Moreover,
we now explicitly take into account the variation of the width with
temperature in the model atmosphere calculations, a variation that is
not simply proportional to $T^{-1/2}$, the expected dependency
according to simple classical calculations (Table~1 shows, for
example, the result of the new Stark broadening calculations for the
strong CII $\lambda$4267 line). The widths are scaled linearly with
electron density, a reasonable approximation that is valid for the
densities of interest here (electron densities of the order of $N_e
\sim 10^{18}$ cm$^{-3}$ are reached only in the deepest layers where
$\tau_R > 10$).

\begin{table}[!tH]
\begin{center}
\vbox{\footnotesize\tabcolsep=3pt
\parbox[c]{124mm}{\baselineskip=10pt
{\smallbf\ \ Table 1.}{\small\  
Stark widths for the strong CII 4267 line (N$_e= 10^{17}$cm$^{-3}$).}}
\begin{tabular}{cc}
T (K) & Width (\AA)  \\
\hline
5,000 & 2.08 \\
10,000 & 1.67 \\
20,000 & 1.37 \\
30,000 & 1.26 \\
50,000 & 1.16 \\
\hline
\end{tabular}
}
\end{center}
\vskip-8mm
\end{table}

Using these state-of-the-art Stark broadening parameters for CII
lines, we next computed a new model atmosphere grid for Hot DQ
stars. This new generation of model atmosphere also include several
improvements over those presented in Dufour et al. (2008). The
numerous modifications made to our code will be reported in details in
a forthcoming publication. Our grid covers a range from $T_{\rm eff}$
= 16,000 to 25,000~K in steps of 1,000~K, from log g= 7.5 to 10.0 in
steps of 0.5 dex, and from log (C/H)= +3.0 to 0.0 in steps of 1.0
dex. This grid has been calculated with a fixed value of C/O = 1.0 for
this exploratory study, a value approximately appropriate for
SDSS~J1153+0056 according to preliminary inquiry. Proper navigation in
the C/O dimension will be done in due time.

We first focus on the simpler objects which do not show sign of
magnetic line splitting (limits of about $\sim$300-400 kG given the
spectral resolution of our observations). Our spectroscopic fitting
procedure relies on the standard nonlinear least-squares method of
Levenberg-Marquardt and is similar to that described in Dufour et
al. (2008). Figure 1 shows an example of our best fit solution for
SDSS~J1153+0056. We refrain, however, from giving final atmospheric
parameters at this point since, in this preliminary study, we
calculated only one grid with a fixed oxygen abundance. Moreover, new
oxygen line Stark widths calculations, which are currently underway,
will also need to be included in replacement of the approximation used
in our grid. As a consequence, it is expected that the atmospheric
parameters that we derive here will change slightly when all the
correct ingredients are put together and the parameter space
explored appropriately. 

Nevertheless, we can already notice significant improvements in the
quality of our fits compared to our first generation of models (see
fig 2 of Dufour et al. 2009 for example). Quantitatively, we also
observe significant difference between the parameters determined in
Dufour et al. (2008) and those found with our new model atmosphere
grid. For instance, we find surface gravities much higher than the log
g $\sim$8 reported in Dufour et al. (2008), with values now in the
vicinity of log g $\sim$9. Such differences are due to a combination
of several factors: better S/N ratio spectroscopic observations,
improved continuum opacities, new Stark broadening parameters and the
presence of large amount of oxygen that was previously unnoticed. It
thus appear that Hot DQ stars are among the most massive white dwarfs
known. Unfortunately, as noted above, there are still more calculations
that remain to be completed before we can provide a more quantitative
assessment of this affirmation. We must, however, remain cautious
about such an interpretation since it is possible that the line
profile appear broader as a result of unresolved components of lines
slightly split by a weak magnetic field. Further high resolution
spectropolarimetric observations, which we hope to obtain soon, should
alleviate this issue.


\begin{figure}[!tH]
\vbox{
\centerline{\psfig{figure=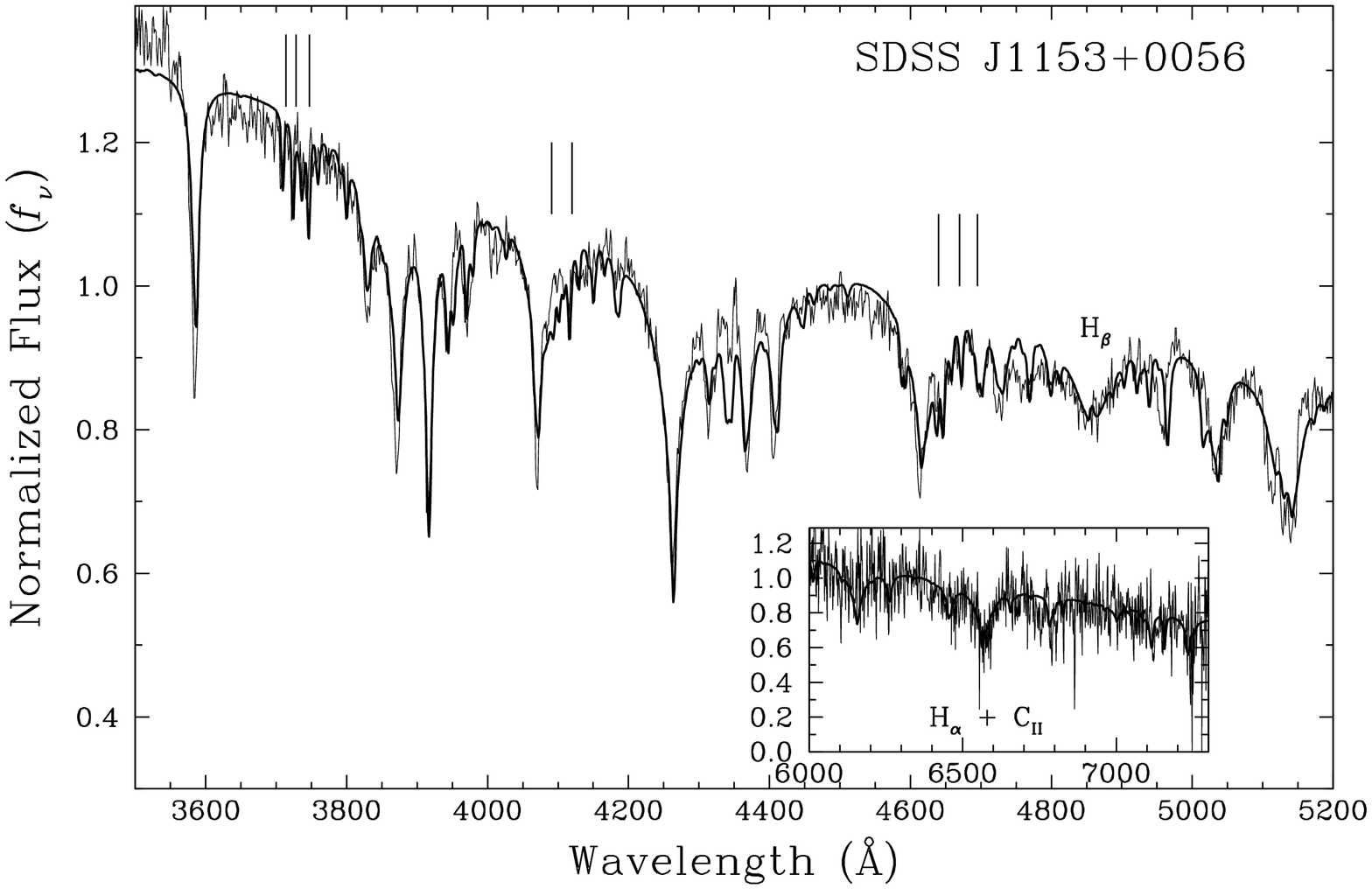,width=100mm,angle=-90,clip=}}
\vspace{1mm}
\captionb{1} {Fit to the carbon lines for the Hot DQ SDSS~J1153+0056
  (oxygen lines, which are not fitted, are indicated by tick marks). The
  thick line is the best solution obtained by fitting the optical
  (MMT) data. The insert shows the H$\alpha$ region (SDSS
  spectroscopic observations). The C/O ratio is fixed to 1 in this
  preliminary analysis. This represent a significant improvement over
  the "first generation" fits of Dufour et al. 2009 (see their figure
  2).}
}
\end{figure}

\sectionb{3}{CONCLUSIONS}

A new generation of model atmosphere including state-of-the-art CII
Stark broadening data is presented for a detailed modeling of Hot DQ
white dwarfs. The new Stark broadening data calculated specifically
for this work will soon be available in the STARK-B database
(http://stark-b.obspm.fr/).

Fits to high S/N spectroscopic data using these new models yield
atmospheric parameters that are significantly different to those
presented in Dufour et al. (2008). As a consequence, it is now
believed that the Hot DQ stars might be among the most massive
isolated white dwarfs that can form from standard stellar evolutionary
channels. However, further calculations are still required before
precise and final atmospheric parameters can be published with
confidence. For instance, OI, OII and CI Stark broadening data also
need to be incorporated in our model atmosphere calculations. 

Future work will also focus on the modelisation of the UV region
(spectroscopic data from HST/COS are now available for 5 Hot DQ, see
Dufour et al. 2010) since, for such hot stars, most of the flux is
emitted in that part of the electromagnetic spectrum. Furthermore,
model atmosphere including magnetic fields with different geometry will
need to be developed for the analysis of the majority of Hot DQ white
dwarfs. As the spectroscopic modeling of the Hot DQ stars gains in
maturity and more accurate atmospheric parameters become available, a
better understanding of the origin and evolution of these strange
stars should soon emerge.

\thanks{P.D. is CRAQ postdoctoral fellow. This work was supported in
  part by the NSERC of Canada and FQRNT Qu\'ebec. It is also supported
  in part by the project 176002 of the Ministry of Education and
  Science of Serbia, by the bilateral cooperation agreement between
  Tunisia (DGRS) and France (CNRS) (project code 09/R 13.03,
  No.22637), by the Paris Observatory and by the Programme National de
  Physique Stellaire (INSU-CNRS).}

\References

\refb Barlow B.N., Dunlap B.H., Rosen R., Clemens J.C. 2008, ApJ, 688,
95

\refb Blagojevic B., Popovic M. V., Konjevic N. 1999, PhyS, 59, 374

\refb Castelli F. 2005, MSAIS, 8, 44

\refb Djenize S., Sreckovic A., Milosavljevic M., Labat O., Platisa M.
1988, ZPhyD, 9, 129

\refb Dufour P., Liebert J., Fontaine G. and Behara N. 2007, Nature,
450, 522

\refb Dufour P., Fontaine G., Liebert J., Schmidt G. D. and Behara
N. 2008, ApJ, 683, 978

\refb Dufour P., Liebert J., Swift B., Fontaine G., and Sukhbold
T. 2009, JPhCS, 172, 012012

\refb Dufour P., Fontaine G., Bergeron P., B\'eland S., Chayer P.,
Williams K.A. and Liebert J. 2010, 17th European White Dwarf Workshop,
AIP Conference Proceedings, 1273, 64

\refb Dufour P., B\'eland S., Fontaine G., Chayer P., Bergeron
P. 2011, ApJ, 733, 19

\refb Dunlap B.H., Barlow B.N., Clemens J.C. 2010, ApJ, 720, 159

\refb Goly A., Weniger S. 1982, JQSRT, 28, 389

\refb Griem H.R, 1974, Spectral line broadening by plasmas, Academic
Press, New York

\refb Larbi-Terzi N., Sahal-Br\'echot S., Ben Nessib N., Dimitrijevic
M.S.  2010, 17th European White Dwarf Workshop, AIP Conference
Proceedings, 1273, 428

\refb Mahmoudi W.F., Ben Nessib B. and Sahal-Br\'echot S. 2004, PhyS,
70, 142

\refb Montgomery M.H., Williams K.A., winget D.E., Dufour P., De
Gennaro S., Liebert J. 2008, ApJ, 678, 51

\refb Peytremann E., 1972, A\&A, 17, 76

\refb Sahal-Br\'echot S. 2010, JPhCS, 257, 012028

\refb Sahal-Br\'echot S., Dimitrijevic M.S. and Ben Nessib N. 2011,
Baltic Astronomy, xx, xxx (this issue)

\refb Sahal-Br\'echot S. and Segre E.R.A., 1971, A\&A, 13, 161

\refb Sreckovic A., Drincic V., Bukvic S., Djenize S. 2000, JPhB, 33, 4873

\end{document}